\documentclass[sigconf]{acmart}
\usepackage{graphicx}
\usepackage{amsmath}
\usepackage{booktabs}
\usepackage{amsfonts}
\usepackage{multirow}[c]
\usepackage{makecell}
\usepackage{subfig}
\usepackage{color}
\usepackage[table]{xcolor}
\usepackage{bm}
\usepackage{epstopdf}
\usepackage{url}
\usepackage[cal=cm]{mathalfa}
\usepackage{balance}
\usepackage{threeparttable}
\usepackage{wrapfig}
\usepackage{tabularx}
\usepackage{array}
\usepackage{enumitem}
\usepackage{appendix}
\usepackage{tikz}
\usepackage{float} 
\usepackage{adjustbox}
\usepackage[linesnumbered,ruled,vlined]{algorithm2e}

\SetKwProg{Stage}{Stage}{}{} 

\setlist[itemize]{leftmargin=*}

\AtBeginDocument{}

\renewcommand\footnotetextcopyrightpermission[1]{}

\begin{document}

\author{
    Shen Wang\textsuperscript{$\dag$}, 
    Yusheng Huang\textsuperscript{$\dag$},
    Ruochen Yang\textsuperscript{$\dag \ast$},
    Shuang Wen\textsuperscript{$\dag$},
    Pengbo Xu\textsuperscript{$\dag$},
    Jiangxia Cao\textsuperscript{$\ddag$}, \\
    Yueyang Liu,
    Kuo Cai,
    Chengcheng Guo,
    Shiyao Wang,
    Xinchen Luo,
    Qiang Luo, \\
    Ruiming Tang,
    Shuang Yang,
    Zhaojie Liu,
    Guorui Zhou,
    Han Li,
    Kun Gai
}
\affiliation{
\institution{
    Kuaishou Technology, Beijing, China
}
\country{
    \{wangshen, huangyusheng, yangruochen, wenshuang, xupengbo, caojiangxia, liuyueyang05, caikuo, guochengcheng03, wangshiyao08, luoxinchen, luoqiang, tangruiming, yangshuang08, zhaotianxing, zhouguorui, lihan08\}@kuaishou.com, kun.gai@qq.com
}
\thanks{\noindent
    $\dag$ Equal Contribution. \\
    $\ddag$ Jiangxia Cao is Corresponding Author.\\
    $\ast$ Ruochen Yang is an internship at Kuaishou. He is now at Institute of Information Engineering, Chinese Academy of Sciences.
}
}

\renewcommand{\shortauthors}{Kuaishou Reco et al.}

\title{OneLive: Dynamically Unified Generative Framework for Live-Streaming Recommendation}

\renewcommand{\shorttitle}{OneLive}

\begin{abstract}

Live-streaming recommender system serves as critical infrastructure that bridges the patterns of real-time interactions between users and authors. 
Similar to traditional industrial recommender systems, live-streaming recommendation also relies on cascade architectures to support large-scale concurrency.
Recent advances in generative recommendation unify the multi-stage recommendation process with Transformer-based architectures, offering improved scalability and higher computational efficiency.
However, the inherent complexity of live-streaming prevents the direct transfer of these methods to live-streaming scenario, where continuously evolving content, limited lifecycles, strict real-time constraints, and heterogeneous multi-objectives introduce unique challenges that invalidate static tokenization and conventional model framework.

To address these issues, we propose \textbf{OneLive}, a dynamically unified generative recommendation framework tailored for live-streaming scenario. OneLive integrates four key components: (i) A \textbf{Dynamic Tokenizer} that continuously encodes evolving real-time live content fused with behavior signal through residual quantization; (ii) A \textbf{Time-Aware Gated Attention} mechanism that explicitly models temporal dynamics for timely decision making; (iii) An efficient decoder-only generative architecture enhanced with \textbf{Sequential MTP} and \textbf{QK Norm} for stable training and accelerated inference; (iv) A \textbf{Unified Multi-Objective Alignment Framework} reinforces policy optimization for personalized preferences.  
Extensive offline experiments demonstrate the outperformance of our model. 
We have deployed OneLive in Kuaishou’s live-streaming recommendation system, where online A/B tests show significant gains on multiple core business indicators, validating its effectiveness and practicality in real-world industrial scenarios.

\end{abstract}

\maketitle

\section{Introduction}

In recent years, industrial recommender systems~\cite{moment, tsstfn} have played a pivotal role across a wide range of online applications, including e-commerce~\cite{din}, content distribution~\cite{moment}, advertising~\cite{gpr} and local services~\cite{oneloc}. Their primary objective is to efficiently and accurately match users' preferences with relevant content under large-scale, high-concurrency real-world conditions, thereby enhancing user experience while maximizing key business metrics.

However, the relentless pursuit of higher accuracy has driven the evolution of recommendation models toward increasingly complex and sophisticated system architectures:
\begin{itemize}
    \item \textbf{Cascade Architecture}: Recommender systems are required to perform personalized matching and ranking over massive candidate pools within an extremely short latency. To meet this demand, the industry has long adopted a cascade architecture as \textbf{\textit{Retrieval - Pre-Ranking - Ranking}}. However, this funnel-shaped framework imposes fundamentally misaligned objectives and constraints across stages. 
    \textbf{\textit{Retrieval}} aims to efficiently narrow down billions of items to a few thousand relevant candidates, prioritizing coverage and diversity under strict computational budgets. 
    \textbf{\textit{Ranking}} focuses on accurately estimating user preference for each candidate, where precision and multi-objective optimization are paramount.
    This misalignment often leads to locally optimal but globally suboptimal results, since the performance ceiling of downstream stages is inherently capped by the quality and diversity of candidates passed from upstream. Once superior items are prematurely filtered out in early stages, an irrecoverable information bottleneck is created, resulting in suboptimal recommendation and low efficiency in model iteration. 
    \item \textbf{Fragmented Computational Graph}: Typical industrial recommender architecture comprise a heterogeneous mix of components, including sparse embedding lookups, sequential modeling and feature interaction modules, which results a highly irregular computational graph. This graph is dominated by memory accesses and data pipeline transfer, which fragment the execution flow. Consequently, the system becomes memory-bandwidth bound rather than compute bound, as the \textbf{Model FLOPS Utilization (MFU)} is often extremely low due to the ratio of effective floating-point operations in time consumption is lower than that of memory access and synchronization.
\end{itemize}

As the Transformer architecture has become the maintain standard in LLM~\cite{qwen3, deepseek-v3}, which has demonstrated consistent gains with scaling, recommender systems are also showing a trend of converging towards th Transformer paradigm~\cite{tiger, hstu, onetrans}. The core value of Transformers lies in scalability and computational efficiency. 
On the one hand, Transformer provides a unified modeling interface, which can organize heterogeneous signals from multiple sources like attribute features and behavior contexts into token sequences, enabling representation learning and interaction modeling within a single framework.
On the other hand, Transformers are built around attention and FFN, whose computation is dominated by large-scale matrix operations, thereby improving MFU through parallelization structure and obtaining benefits by scaling model capacity and training data.

\begin{figure}[t!]
\begin{center}
\includegraphics[width=8.5cm]{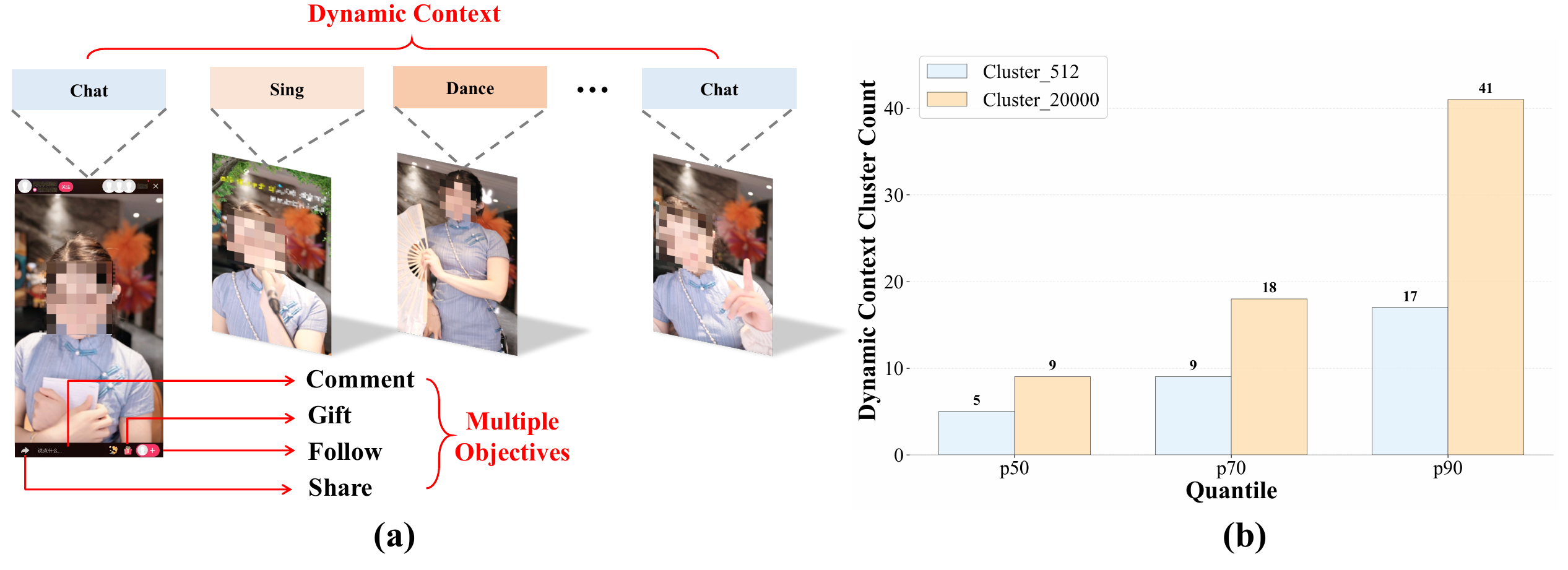}
\vspace{-0.3cm}
\caption{(a) Over a live-streaming lifecycle, the author exhibits diverse content types, accompanied by multi-objective user interactions. (b) Clustering 30-second segments across entire live-streamings and report the mean number of clusters to quantify continuous content dynamics.
}

\vspace{-0.5cm}
\label{fig:motivation}
\end{center}
\end{figure}

Following this paradigm, OneRec~\cite{onerec} and a series of related \textbf{Generative Recommendation (GR)} methods~\cite{onerec-v2, onesug, oneloc, onesearch} have been proposed. These large recommendation models share a common theme that they advocate using a single and unified model architecture to accomplish end-to-end recommendation, together with preference alignmnet and engineering efforts for high throughput deployment.
Concretely speaking, they tokenize business entities via residual quantization, perform autoregressive, session-wise generation with a unified Transformer backbone, and further align the model to fine-grained preference signals through reinforcement learning. 
By removing the systemic overhead of traditional multi-stage cascaded pipelines, this line of work achieves a more favorable cost structure and stronger objective consistency, and has consequently attracted substantial interest from industry~\cite{gpr, mmq-v2}.

However, despite their general effectiveness, such methods cannot be directly applied to live-streaming recommendation without careful adaptation, owing to the inherent complexity of the live-streaming scenario:
\begin{itemize}
    \item \textbf{Dynamic Content}: Unlike short videos or image-text contents that are fixed after users' upload, live-streaming content evolves continuously throughout the broadcast. Correspondingly, author activities, real-time user feedback and the overall interactive atmosphere all undergo substantial changes over time.
    For example, an author may frequently switch among chat, sing and dance like Figure \ref{fig:motivation}(a). 
    This dynamic nature challenges the standard tokenization paradigm in GR, where an item’s static content can be encoded once into a stable representation for repeated use. 
    In contrast, such a static tokenization paradigm fails to accommodate the dynamic nature of live-streaming, 
    as the real-time content clustering statistics of the live-streaming shown in Figure \ref{fig:motivation}(b).
    This necessitates dynamic modeling and tokenization with real-time content understanding and update to track live-streaming content drift.

    \item \textbf{Limited Lifecycle}: Live-streaming exhibit an inherent lifecycle constraint, where the full progression from initiation, growth, peak, decline to termination is typically confined within a short time window. 
    This results in a strictly time-constrained candidate pool, where only ongoing live-streamings are eligible for recommendation, rather than the persistent content inventory in conventional platforms. 
    A live-streaming must be promoted within its golden exposure window to realize its full value. 
    Consequently, live-streaming GR must prioritize the validity of generated candidates throughout their entire lifecycle.

    \item \textbf{Real-Time Response Constraint}: Live-streaming recommendation operates in a high concurrent and critical latency online environment, where the candidate pool evolves rapidly and user queries arrive at extremely high frequency. 
    This imposes stringent requirements on the live-streaming GR model, which must maintain competitive prediction accuracy while achieving efficient inference with low latency and sustained throughput.

    \item \textbf{Multiple Objectives}: Live-streaming recommendation is a multi-objective task with rich user feedback signals, including click, share, follow and gift in Figure \ref{fig:motivation}(a). 
    Users also exhibit distinct preferences across various forms of engagement and consumption behaviors, resulting in pronounced heterogeneity.
    Therefore, preference alignment over multi-objective signals is essential for live-streaming GR, requiring the integration of diverse supervision signals into a unified generative training pipeline.

\end{itemize}

\begin{figure*}[ht!]
\begin{center}
\includegraphics[width=17cm]{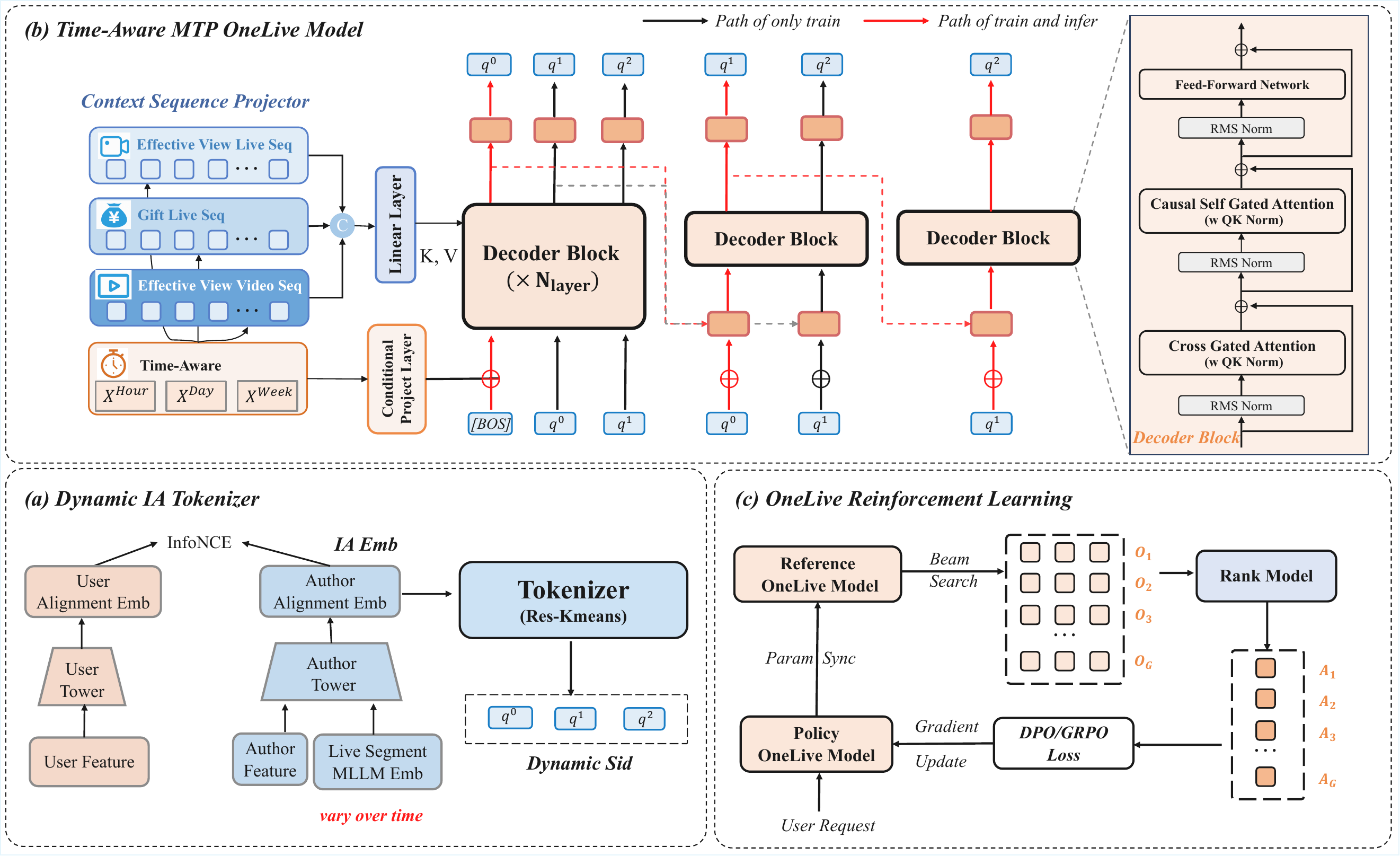}
\vspace{-0.3cm}
\caption{Overall framework of OneLive.}
\vspace{-0.5cm}
\label{fig:model}
\end{center}
\end{figure*}

To this end, we propose \textbf{OneLive}, a dynamically unified generative framework for live-streaming recommendation. Composed of a dynamic tokenizer, an efficient generative architecture, and a multi-objective alignment module, OneLive effectively addresses the unique challenges of live-streaming scenarios. To the best of our knowledge, OneLive is the first unified generative recommendation framework that has been successfully deployed on a large-scale, real-world live-streaming platform. We summarize the main contributions of this work as follows:

\begin{itemize}
    \item We propose a novel \textbf{Dynamic Tokenizer} tailored to the inherently evolving nature of live-streaming content. By capturing real-time content information and fusing it with user behavioral signals, together with residual quantization, our method enables effective author representation learning and compression driven by online recommendation signals.

    \item We propose a \textbf{Time-Aware Gated Attention} mechanism to model the strict timing constraints imposed by the limited lifecycle of live-streaming. It explicitly incorporates the evolving patterns and dynamic user feedback into interaction modeling and decision making, thereby meeting the stringent timeliness requirements of live-streaming distribution.
    
    \item We propose a \textbf{Sequential Multi-Token Prediction (MTP)} mechanism to accelerate inference in the decoder-only GR architecture, and further introduce \textbf{QK Normalization} as a key component to ensure stable training and reliable online serving.
    
    \item We propose a \textbf{Unified Multi-Objective Alignment Framework} powered by reinforcement learning with an ensemble ranking reward model. This framework explicitly fuses personalized multi-objective and aligns heterogeneous user preferences to accommodate diverse behaviors in live-streaming scenarios.
    
    \item We conduct extensive offline experiments to validate our framework and individual modules. We have deployed OneLive on Kuaishou’s live-streaming recommendation system, and large-scale online A/B tests show significant and consistent improvements on core business metrics, verifying its practical value in real industrial scenarios.
\end{itemize}

\section{Methodology}

In this section, we propose OneLive, an end-to-end generative framework for live-streaming recommendation. The overall architecture of our framework is illustrated in Figure \ref{fig:model}.

\subsection{Dynamic Tokenizer} \label{sec:tokenizer}
As commonly practiced in existing GR methods~\cite{tiger, onerec}, the generation of item embeddings followed by semantic quantization constitutes a core operation in tokenization pipelines. 
This two-step process balances the commonalities and distinct characteristics among candidate items, and effectively preserves essential information while controlling vocabulary size. 
Conventionally, this pipeline focuses on items with fixed content after upload, 
where it first captures the comprehensive content of such items to generate item embeddings, and subsequently quantizes these item embeddings into a set of multi-level codes, referred to as Semantic IDs (SIDs), via methods such as RQ-VAE~\cite{rq-vae} or Res-Kmeans~\cite{qarm}. 
Some alternative approaches~\cite{letter, mmq-v2} further incorporate behavioral information as collaborative signals into content understanding, thereby enhancing the quality of item embedding generation.

However, the dual dynamics of live-streaming scenarios pose non-trivial challenges to the direct application of these conventional pipelines: (i) \textit{\textbf{Dynamic streaming content}}: Authors engage in diverse activities (\textit{e.g.}, chatting, singing and dancing) over time, causing real-time changes in live-streaming content. (ii) \textit{\textbf{Dynamic user behavior}}: Temporal changes in live-streaming content directly affect users' preferences commenting, gifting and other interactions. Therefore, it is crucial for a tokenizer to address the dual dynamics issue. 
Consequently, we propose a unified dynamic tokenizer in OneLive, which jointly models real-time live-streaming content and users' immediate behavioral feedback during the item embedding generation stage.

\begin{table*}[t!]
  \caption{The Code Utilization Rate (UR) and Collision Rate (CR) under diverse codebook settings. The embeddings are collected from about 4 million authors. The results of best combination are boldfaced. Metrics definitions are provided in Appendix \ref{app:codebook}.}
  \vspace{-0.3cm}
  \label{tab:code}
    \begin{tabular}{c|c|ccccccc}
    \toprule
    \multicolumn{1}{c}{\multirow{2.5}{*}{\textbf{Type}}} & \multicolumn{1}{c}{\multirow{2.5}{*}{\textbf{Method}}} & \multirow{2.5}{*}{\textbf{Size}} & \multicolumn{4}{c}{\textbf{UR $\uparrow$}} & \multicolumn{2}{c}{\textbf{CR $\downarrow$}} \\
    \cmidrule(r){4-7} \cmidrule(r){8-9} \multicolumn{1}{c}{ } & \multicolumn{1}{c}{ } & & \textbf{$\bm{L_0}$} & \textbf{$\bm{L_1}$} & \textbf{$\bm{L_2}$} & \textbf{$\bm{L_0}\times\bm{L_1}$} &  \textbf{SID} & \textbf{Author} \\
    
    \midrule
    
    \multirow{2.5}{*}{\textbf{MLLM}} & \multirow{2.5}{*}{Res-Kmeans} & (512, 512, 512) & 100.00\% & 99.61\% & 97.66\% & 53.83\% & 28.10\% & 63.83\% \\
    \cmidrule(r){3-3} &  & (8192, 8192, 8192) & 88.99\% & 63.93\% & 53.10\% & 2.34\% & 3.56\% & 9.47\% \\
    
    \midrule 
    
    \multirow{5}{*}{\textbf{IA}} & \multirow{2.5}{*}{RQ-VAE} & (512, 512, 512) & 100.00\% & 100.00\% & 100.00\% & 74.41\% & 15.76\% & 39.76\% \\
    \cmidrule(r){3-3} &  & (8192, 8192, 8192) & 100.00\% & 100.00\% & 100.00\% & 1.16\% & 8.54\% & 22.78\% \\
    \cmidrule(r){2-3} & \multirow{2.5}{*}{\textbf{Res-Kmeans}} & (512, 512, 512) & 100.00\% & 100.00\% & 100.00\% & 93.98\% & 9.72\% & 23.03\% \\
    \cmidrule(r){3-3} &  & \textbf{(8192, 8192, 8192)} & \textbf{100.00\%} & \textbf{100.00\%} & \textbf{100.00\%} & \textbf{4.50\%} & \textbf{0.66\%} & \textbf{1.76\%} \\
    
    \bottomrule
  \end{tabular}
  \vspace{-0.3cm}
\end{table*}


\subsubsection{\textbf{Semantic and Collaborative Dynamic Alignment}}
Live-streaming real-time content is multimodal, encompassing visual, auditory, and textual information. 
However, content-only embeddings lack collaborative signals and suffer from low discriminability, such that semantically similar clips may yield nearly identical representations and thus fail to distinguish different streamers.
Naive embedding fusion with author side information fails to capture evolving popularity or cohort-specific preferences~\cite{mmq-v2}. 
Alternatively, directly injecting collaborative signals into MLLM pre-training~\cite{letter, eager} is ill-suited for live streaming.
These approaches rely on static, pre-collected behavior data, while user-author collaborative patterns evolve rapidly and unpredictably in real-time, which cannot be captured by offline pretraining.

To address this dilemma, we introduce a two-stage paradigm consisting of \textit{dynamic content understanding} and subsequent \textit{real-time collaborative post-alignment}, which jointly generates the final item embedding for real-time live-streaming segments.

\textit{Dynamic Content Understanding.}
To model cross-modal live-streaming content, we employ a MLLM to generate representation embeddings. Given the trade-off between contextual integrity and online latency constraints, we balance these demands by adopting a 30-second sliding window to dynamically update live-streaming content embeddings~\cite{moment} in practice.
We leverage a large LLM to curate a dataset of content clips, and further fine-tune a lightweight MLLM through knowledge distillation~\cite{larm}, which continuously ingests real-time segmented content to generate live-streaming dynamic embeddings.
\textit{Real-time Collaborative Post-alignment.}
Building upon the dynamic content embeddings obtained above, we further conduct real-time post-alignment with collaborative signals using a dual-tower architecture. This stage explicitly integrates instant user interaction feedback (e.g., commenting, gifting, clicking) to complement semantics-only representations, thereby alleviating discriminability issues and adapting to the highly dynamic user-author collaboration patterns.

Specifically, in the author tower, we fuse the author's static attributes $x^{AId} \in \mathbb{R} ^d$ with dynamic content features. The latter includes real-time content embedding $x^{MLLM}_{30s} \in \mathbb{R} ^d$ derived from the most recent 30-second window to display instant status, and long-term average pooling embedding  $x^{MLLM}_{pooling} \in \mathbb{R} ^d$ which reflects author's thematic commonalities. We employ a gating mechanism for representation fusion on the author side:
\begin{equation}
\label{eq:MLLM_fusion}
    x^{MLLM} = \text{MLP}(x^{MLLM}_{30s} \oplus x^{MLLM}_{pooling}),
\end{equation}
\begin{equation}
\label{eq:author_fusion}
    x^{Author} = \lambda x^{AId} + (1 - \lambda) x^{MLLM},
\end{equation}
where $\lambda$ is a trainable weight to balance heterogeneous inputs.

The user's attributes $x^{UId} \in \mathbb{R} ^d$ are processed through the user tower, and then we peform user-author alignment based on the interaction samples in the data streaming:
\begin{equation}
\label{eq:user_fusion}
    x^{User} = \text{MLP}(x^{UId}),
\end{equation}
\begin{equation}
\label{eq:fm}
    \mathcal{L}_{Alignment} := \text{In-Batch-Softmax}(x^{User}, x^{Author}).
\end{equation}

The output embedding of the author tower $x^{Author}$ preserves author attributes and real-time content generalization via gated fusion, and incorporate collaborative signals through dual-tower interaction supervision. 
We obtain the final semantically and collaboratively aligned item embedding for live-streaming GR, termed IA Embedding $x^{IA} := x^{Author}$, for downstream quantification tasks.

\subsubsection{\textbf{Residual Quantization}}

To pursue better scalability and more explicit hierarchical semantics, we opt for Res-Kmeans over RQ-VAE for residual quantization. Specifically, the IA Embedding is quantized in a coarse-to-fine manner, where K-means is applied iteratively to the multi-level residuals $\mathcal{R}^{l}$ to heuristically construct the codebook at each layer:
\begin{equation}
\label{eq:codebook}
    q^{l} = \arg\min_k \lVert c_k^l - x^l \rVert,
\end{equation}
where $c^l_k \in \mathcal{C}^{l} = \text{Kmeans}(x^{l}, N_{l})$ is the codebook derived from $N_{l}$ centroids obtained through clustering, and $x^{l+1} = x^{l} - c^{l}_{q^l}$ is the residual of $l$ layer as the input for the $l+1$ layer. By iterating to obtain the quantization code of each layer $q^{l}$ for $T$ times, we can get the SID corresponding to the author in real time $\{q^0, q^1, \dots, q^T\}$. 

We compare the quantization quality under different settings, with results reported in Table~\ref{tab:code}. IA Embeddings achieve higher codebook utilization and lower code collision rates than MLLM Embeddings, owing to refinement via collaborative post-alignment and supervision from user-item interaction signals. Similarly, Res-Kmeans quantization outperforms RQ-VAE with improved codebook utilization and fewer collisions under the same settings, and increasing the codebook size effectively alleviates code collisions. We thus employ Res-Kmeans with $T=3$ layers and codebook size $N_l=8192$ for semantic quantization.

\subsection{Generative Recommender} \label{sec:dec}

OneLive adopts a lightweight and time-aware architecture. Built upon a standard decoder-only backbone, we model temporal dynamics via a gated attention mechanism, improve inference efficiency through sequential multi-token prediction, and enhance training stability with QK Norm.

\subsubsection{\textbf{Basic Decoder-Only Architecture.}}
In industrial live-streaming recommendation, modeling user preferences over authors relies on integrating diverse behavior sequences (e.g., live-streaming effective viewing $S^{Eff}$, gifting $S^{Gift}$, short-video viewing $S^{Video}$) to capture fine-grained interests. However, encoding each sequence respectively with Encoder-Decoder architectures like T5~\cite{t5}, TIGER~\cite{tiger} incurs substantial overhead.
The core issues is that the sequences are numerous and lengthy, with total length far exceeding the target streamer SID (\textit{e.g.}, $3{\times}500 \gg 3$, where $n=3, l=500$ denote the number and per-sequence length). 
As a result, vast majority of computation goes to the encoder stage, wasted on redundant or noisy interactions in long sequences rather than target SID generation.

Inspired by~\cite{onerec-v2}, we adopt a lazy decoder-only architecture that avoids explicit encoding of lengthy user sequences, while maintaining model capacity comparable to conventional encoder-decoder structures for complex user-author interactions.

Specifically, we project user's multiple historical sequences into an embedding space, which serves as key-value contexts for generation:
\begin{equation}
    S = \{S^{Eff}, S^{Gift}, S^{Video}, \dots\}_n ,
\end{equation}
\begin{equation}
\label{eq:feature_seq}
    x_i = \text{Concat}(x_i^{\text{AId}}, x_i^{\text{IA}}, \dots) = \text{Emb}\left(\mathcal{S}[i]\right) \in \mathbb{R}^d,
\end{equation}
\begin{equation}
\label{eq:seq}
    E = \text{MLP}(\{x_0, x_1, \dots, x_{|S|}\}),
\end{equation}
where $E \in \mathbb{R} ^ {|S|\times d}$ is the encoding of user behavioral signals, and $|S| = n\times l$ is the length of concatenated sequence.

The Decoder-Only architecture is composed of $L$ stacked transformer blocks, with three main components: cross-attention, self-attention and feed-forward network (FFN). 
Formally, the computation of the $l$-th layer is:
\begin{equation}
\label{eq:crossattn}
    h^l = h^{l-1} + \text{CrossAttn}(q=\text{RMSNorm}(h^{l-1}), k=E, v=E),
\end{equation}
\begin{equation}
\label{eq:selfattn}
    h^l = h^{l} + \text{SelfAttn}(\text{RMSNorm}(h^{l})),
\end{equation}
\begin{equation}
\label{eq:ffn}
    h^l = h^{l} + \text{FFN}(\text{RMSNorm}(h^{l})),
\end{equation}
where $h^l$ denotes the hidden state of each layer. For the first layer, $h^0 = \text{Emb}(\{[\text{BOS}], q^0, q^1\}) = \{x_{[\text{BOS}]}, x_{q^0}, x_{q^1}\} \in \mathbb{R} ^ {3\times d}$ under the teacher-forcing training paradigm.

\subsubsection{\textbf{Temporal Dynamics Conditioning.}}
Live-streaming recommendation is a typical time-constrained distribution scenario that aligns with the inherent limited lifecycle of live-streamings, where only ongoing live broadcasts from online authors are eligible for recommendation and system distribution.
To address this strict temporal constraint, we model temporal dynamics from three perspectives:
\textit{historical sequence temporal perception}, \textit{generation anchor temporal perception}, and \textit{attention-gated temporal perception}.

\textit{Historical Sequence Temporal Perception}.
We inject multi-granular time biases (e.g., hour, day, week) associated with each historical item $x_i$ (Eq.~\ref{eq:feature_seq}) as temporal signals to construct time-aware sequential representations:
\begin{equation}
    x_{i\_ta} = x_i + \text{MLP}\left(\text{Concat}(x_i^{\text{Hour}}, x_i^{\text{Day}}, x_i^{\text{Week}})\right),
\end{equation}
which are further mapped into the time-aware context embedding $E_{ta}$ following Eq.~\ref{eq:seq}.

\textit{Generation Anchor Temporal Perception}.
As the $[\text{BOS}]$ token is the autoregressive generation anchor in our lazy decoder-only framework, we augment $[\text{BOS}]$ with multi-granular temporal features that mark the instantaneous query time of the user, enabling time-aware conditional generation:
\begin{equation}
    x_{[\text{BOS}]\_ta} = x_{[\text{BOS}]} + \text{MLP}\left(\text{Concat}(x_q^{\text{Hour}}, x_q^{\text{Day}}, x_q^{\text{Week}})\right),
\end{equation}
where $x_q^{*}$ denotes the embedding of temporal features of the query moment.

\textit{Attention-Gated Temporal Perception}.
We equip each SID generation step with adaptive context importance selection by introducing gated attention~\cite{gated-attention} into the decoder. This mechanism dynamically modulates attention weights over both user historical encoding and the preceding decoded sequence via learnable gating scores. It is applied to both cross-attention (Eq.~\ref{eq:crossattn}) and self-attention (Eq.~\ref{eq:selfattn}):
\begin{equation}
    \text{Score}(X) = \sigma(XW_{\theta}),
\end{equation}
\begin{equation}
    O = \left(\text{MultiHeadAttn}(XW_Q, X'W_K, X'W_V) \odot \text{Score}(X)\right)W_O,
\end{equation}
where $X'=E$ for cross-attention and $X'=X$ for self-attention, $W_{\theta} \in \mathbb{R}^{d \times d}$ denotes the gating projection parameter, and $\sigma(\cdot)$ is the Sigmoid activation.
The gating scores are applied via element-wise multiplication after multi-head attention concatenation, allowing the model to adaptively emphasize temporally relevant context.

\subsubsection{\textbf{Sequential Multi-Token Prediction.}}

The standard autoregressive decoder is trained via the next-token prediction (NTP) mechanism, where cross-entropy loss is adopted with teacher-forced ground-truth SID codes $\{q^0,q^1,q^2\}$. The corresponding loss function is formally defined as follows:
\begin{equation}
    \mathcal{L}_{NTP} := - \sum^{2}_{i=0} \log p(q^i|[\text{BOS}], q^{< i}, E),
\end{equation}
\begin{equation}
    p(q^i|[\text{BOS}], q^{< i}, E) = \text{Softmax} ( \text{MLP} (h^L_i) ).
\end{equation}

During the inference phase, beam search is utilized to perform greedy autoregressive generation. The predicted next author’s SID $p(Q \mid E)$ is formally formulated as
\begin{equation}
    p(Q|E) = p(q^0|[\text{BOS}], E) \cdot p(q^1|[\text{BOS}], q^0, E) \cdot p(q^2|[\text{BOS}], q^{0:1}, E).
\end{equation}

A critical limitation of the aforementioned standard inference approach is inefficiency: in practice, upon receiving a user request, a set of $q^0$ are generated using the given [BOS] token, and so forth—subsequently feeding the inferred set of each preceding SID token into the decoder to generate the next, i.e., $q^1$ from $q^0$ and then $q^2$ from $q^1$. As a result, the inference batch size for $q^0$ is relatively small, while the inference batch sizes for $q^1$ and $q^2$ would be much larger—this computational overhead expands rapidly when a large beam size is adopted. Meanwhile, we observe that SID1 and SID2 are residual components of SID0, and their prediction tasks rely on the conditional probability of the already inferred preceding SIDs, making these two subtasks relatively straightforward. 

To address the aforementioned computational bottleneck while leveraging this key observation, we propose a \textbf{Sequential Multi-Token Prediction (Sequential MTP)} mechanism inspired by ~\cite{deepseek-v3}. As shown in Figure \ref{fig:model}(b), our model have three sequentially connected decoders with a shared SID embedding layer. The first MTP module, referred to as the main-decoder, retains the full $L$ layers transformer blocks similar to the standard setting. Subsequent MTP modules adopt a lightweight single-layer decoder block,
which takes as input the concatenation of the previous MTP module’s hidden states and the embedding of the immediately preceding SID:
\begin{equation}
    h_i^{k} = \text{MLP}(\text{Concat}(\{h_i^{' k-1}, \text{Emb}(q^{i})\})),
\end{equation}
\begin{equation}
    h_{i:2}^{' k} = \text{MTP}_{k}(h_{i:2}^{k}).
\end{equation}
To reuse the KV cache generated during $q^0$ inference, the decoder-lite module shares the parameters of the main-decoder’s first layer (except for the FFN component in Eq.\ref{eq:ffn}). The final training task is the weighted fusion of losses from multiple MTP modules:
\begin{equation}
    \mathcal{L}_{MTP} := w_0\mathcal{L}_{\text{MTP}_{main}} + w_1\mathcal{L}_{\text{MTP}_{1}} + w_2\mathcal{L}_{\text{MTP}_{2}},
\end{equation}
\begin{equation}
    \mathcal{L}_{\text{MTP}_{k}} := -\text{log} \ p(q^{k:2}|[\text{BOS}], q^{<2}, E),
\end{equation}
where $w_k$ is smaller for subsequent MTP modules than for the main-decoder.

During inference, we use the main decoder to infer $q^0$, followed by lite decoder for $q^1$ and $q^2$. 
Owing to parameter sharing between modules allows the fully utilization of KV cache for autoregressive generation, 
our Sequential MTP mechanism is expected to achieve higher inference efficiency than the standard decoder-only.

\subsubsection{\textbf{QK Norm Optimization.}}

\begin{figure}[t!]
    \centering
    \subfloat[Trend comparison of loss.\label{fig:loss_line}]{
        \includegraphics[width=8.5cm]{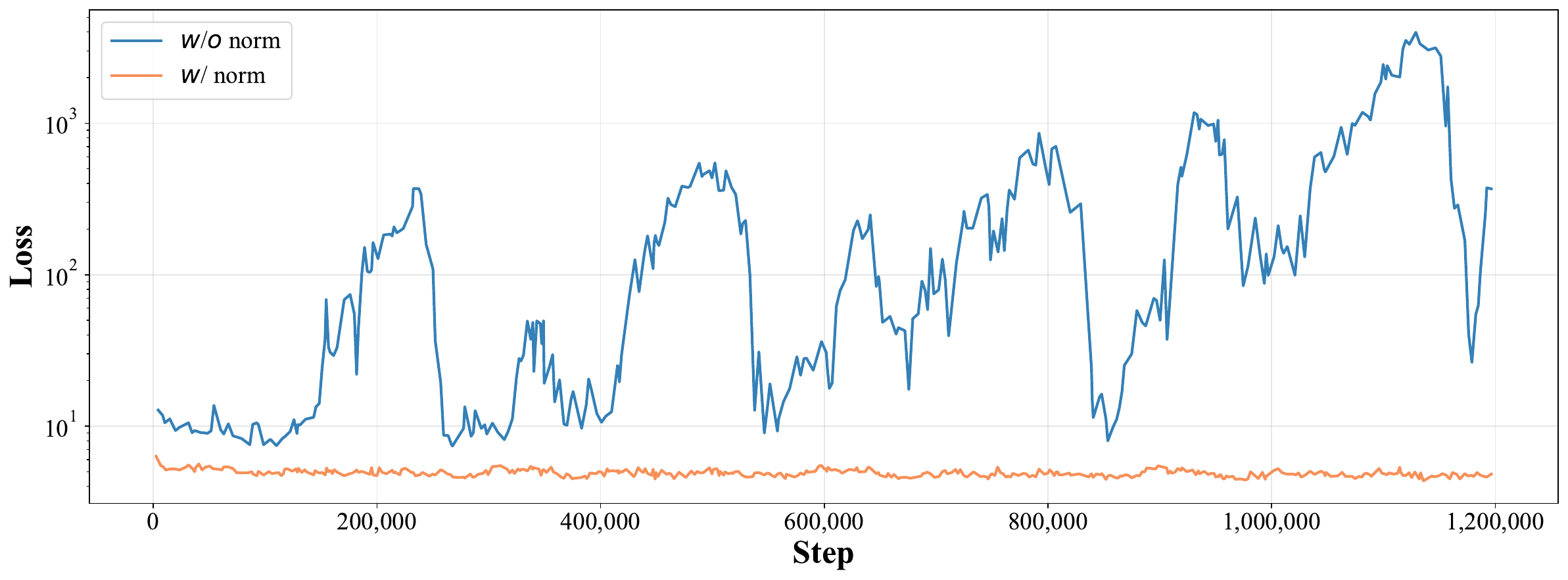}
    }
    \quad
    \subfloat[Trend comparison of max QK logits.\label{fig:qk_line}]{
        \includegraphics[width=8.5cm]{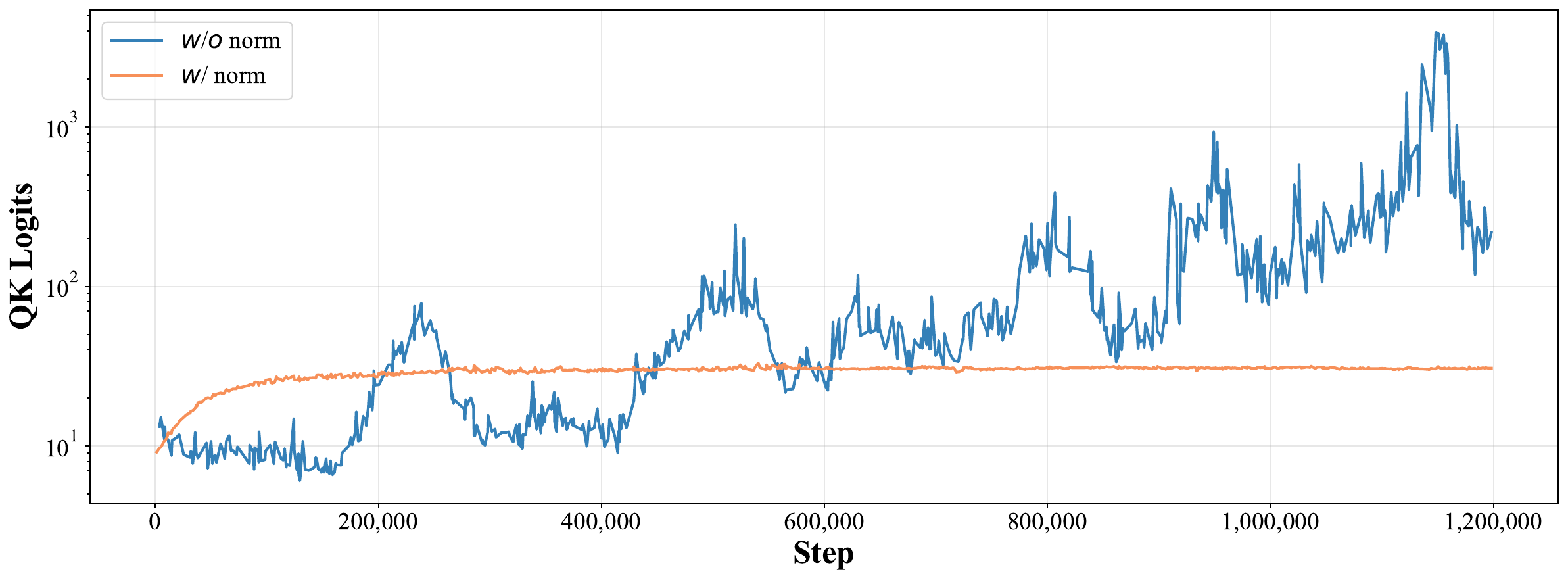}
    }
    \vspace{-0.1cm}
    \caption{The influence of QK Norm on training stability, which  prevents the explosion of loss and max QK logits.}
    \vspace{-0.3cm}
    \label{fig:norm}
\end{figure}

Nevertheless, when further increasing model depth, we observe training instability, as the loss fluctuation shown Figure \ref{fig:norm}(a). 
After prolonged training, the loss occasionally spikes and may fail to recover. 
Diagnostics indicate that the QK logits $QK^T$ in deep attention layers grows substantially over training, exhibiting logit explosion. 
Since we implement bf16 mixed-precision training, the effective numerical precision and dynamic range of the QK logits are limited, with the resulting numerical errors and scale drift can be amplified through deep stacking, progressively expanding the logits~\cite{trainingfails}. 
Once fed into the Softmax, the attention distribution becomes approaching a one-hot vector, which markedly attenuates gradients and can ultimately collapse training.

To improve training stability, we introduce QK Norm~\cite{qknorm}. 
Specifically,  before computing the attention QK logits, we apply RMSNorm to the Query and Key vectors to suppress scale drift across depth and throughout training, making the resulting logits closer to a scale-insensitive similarity measure:
\begin{equation}
    \text{Attn}(Q,K,V) = \text{Softmax}(\frac{\text{RMSNorm}(Q) \text{RMSNorm}(K)^T}{\sqrt{d}})V.
\end{equation}
This normalization effectively constrains the magnitude of the logits and reduces the risk of softmax saturation caused by excessively large values, thereby improving training stability. 
As shown in Figure, with QK Norm enabled, the occasional loss spikes observed during prolonged training disappear.

\subsection{Reinforcement Learning} \label{sec:rl}

Training OneLive solely via next author prediction on offline logs amounts to behavior cloning, which optimizes the model to mimic historical decision patterns of the online serving system. 
Although such a paradigm maintains consistency with past policies, it inherently restricts the model from breaking through the performance ceiling of the deployed system~\cite{onerec}. 
To overcome this limitation, we introduce an explicit reward model and conduct on-policy reinforcement learning, which directly optimizes the generation policy in the generation space using reward signals that reflect users' potential preferences. 
Crucially, the RL objective drives the model to concentrate its generative probability on high-reward candidates, so that the top item set generated by beam search becomes consistent with users' potentially preferred rankings, directly improving the quality of the head generation results at inference time.

\subsubsection{\textbf{Reward Model.}}
The reward for each generated candidate must reflect holistic alignment with users' multi-objective preferences in live-streaming scenarios, including diverse feedback signals such as click, long-view and gift. 
As emphasized in our motivation, live-streaming recommendation exhibits strong user heterogeneity in engagement behaviors, demanding flexible and user-adaptive integration of multiple objectives rather than rigid heuristics. 
A naive baseline is to fuse multiple task-specific scores (e.g., CTR, LVTR, GTR) predicted by a ranking model using fixed handcrafted weights to construct the reward~\cite{onesug, onesearch}. 
Such heuristic fusion has two critical drawbacks, in that fixed weights cannot adapt to user-specific preference heterogeneity and the overall scheme lacks flexibility for online tuning.

To address these issues and enable multi-objective alignment for generative training, we employ the unified multi-objective ensemble ranking model Pantheon~\cite{ltr} as our reward model. 
This model is optimized via a standard additive weighting mechanism to produce a single unified ranking score that inherently balances diverse objectives while adapting to user specific preference heterogeneity:
\begin{equation}
    Score = \text{Pantheon}(\text{User}, \text{Author}),
\end{equation}
\begin{equation}
    \mathcal{L}^{XTR}_{Pantheon} := y^{XTR}\text{log}\ (Score) + (1-y^{XTR})\text{log} \ (1-Score),
\end{equation}
\begin{equation}
    \mathcal{L}_{Pantheon} := \sum^{CTR,LVTR,GTR,\dots}_{XTR} w^{XTR}\mathcal{L}^{XTR}_{Pantheon}.
\end{equation}

We directly use the multi-objective ensemble score $r := Score$ as the reward of each candidate author for corresponding user.

\begin{table*}[ht!]
  \caption{The overall performance comparison of different models on live-streaming offline dataset. The best results are \textbf{boldfaced} and the second-best results are \underline{underlined}, with the relative improvements denoted as \textit{Imprv.}$\uparrow$.}
  \vspace{-0.2cm}
  \label{tab:offline}
  \begin{tabular}{cccccccccc}
    \toprule
    \multicolumn{2}{c}{\multirow{2.5}{*}{\textbf{Models}}} & \multicolumn{4}{c}{\textbf{LongView}} & \multicolumn{4}{c}{\textbf{Click}} \\
    \cmidrule(r){3-6} \cmidrule(r){7-10} & & \textbf{HR@64} & \textbf{MRR@64} & \textbf{HR@128} & \textbf{MRR@128} & \textbf{HR@64} & \textbf{MRR@64} & \textbf{HR@128} & \textbf{MRR@128} \\ 
    \midrule
    \multirow{3}{*}{\textbf{Traditional}} & \multicolumn{1}{|l}{SASRec} & 0.2525 & 0.0665 & 0.3047 & 0.0671 & 0.2394 & 0.0623 & 0.2924 & 0.0629 \\
    & \multicolumn{1}{|l}{KuaiFormer} & 0.3181 & 0.0907 & 0.3762 & 0.0913 & 0.3145 & 0.0890 & 0.3750 & 0.0897 \\
    & \multicolumn{1}{|l}{GNN} & 0.4720 & 0.1548 & 0.5465 & 0.1556 & 0.4565 & 0.1469 & 0.5319 & 0.1478 \\
    \midrule
    \multirow{2}{*}{\textbf{Generative}} & \multicolumn{1}{|l}{TIGER} & 0.2847 & 0.0795 & 0.3395 & 0.0801 & 0.2877 & 0.0785 & 0.3442 & 0.0791 \\
    & \multicolumn{1}{|l}{OneRec} & \underline{0.5967} & \underline{0.2836} & \underline{0.6347} & \underline{0.2840} & \underline{0.5802} & \underline{0.2667} & \underline{0.6209} & \underline{0.2671} \\
    \midrule
    \midrule
    \multirow{2}{*}{\textbf{Ours}} & \multicolumn{1}{|l}{\textbf{OneLive}} & \textbf{0.6887} & \textbf{0.3213} & \textbf{0.7388} & \textbf{0.3218} & \textbf{0.6720} & \textbf{0.3046} & \textbf{0.7246} & \textbf{0.3052} \\
    & \multicolumn{1}{|l}{\textit{\textbf{Imprv. $\uparrow$}}} & \textbf{15.42\%} & \textbf{13.29\%} & \textbf{16.40\%} & \textbf{13.31\%} & \textbf{15.82\%} & \textbf{14.21\%} & \textbf{16.70\%} & \textbf{14.26\%} \\
    \bottomrule
  \end{tabular}
  \vspace{-0.3cm}
\end{table*}

\subsubsection{\textbf{Optimize Strategy.}} 

We explore different preference optimization strategies, including Direct Preference Optimization (DPO)~\cite{dpo} and Group Relative Policy Optimization (GRPO)~\cite{deepseek-r1}. In both cases, a set of response $\{o_1, o_2, \dots, o_G\}$ for each query $q$ is generated through a reference model, which periodically synchronizes parameters from the offline streaming training policy model. The candidate authors in response are then scored by the reward model to obtain reward values $\{r_1, r_2, \dots, r_G\}$, and then the advantage scores are obtained via normalization $A_i = \frac{r_i - \text{mean}(r)}{\text{std}(r)}$. Subsequently, we perform policy optimization:
\begin{itemize}
    \item DPO: We choose the authors with highest advantage as the positive sample $o_{pos}$, and the lowest one as the negative sample $o_{neg}$. Based on these samples, the policy model is optimized to align user preferences:
    \begin{equation}
        \mathcal{L}_{\mathrm{DPO}} := -\sigma(\lambda\log\frac{\pi_{\theta}(o_{p o s}|q)}{\pi_{\theta r e f}(o_{p o s}|q)}-\lambda\log\frac{\pi_{\theta}(o_{n e g}|q)}{\pi_{\theta r e f}(o_{n e g}|q)}) .
    \end{equation}
    \item GRPO: We use the relative advantages of candidate authors to optimize the policy model, enabling it to learn better responses:
    \begin{equation}
        \mathcal{L}_{\mathrm{GRPO}} := \frac{1}{G} \sum^G_{i=1}\text{min}(\frac{\pi_{\theta}(o_{i}|q)}{\pi_{\theta ref}(o_{i}|q)}A_i, \text{clip}(\frac{\pi_{\theta}(o_{i}|q)}{\pi_{\theta ref}(o_{i}|q)}, 1-\epsilon, 1+\epsilon)A_i).
    \end{equation}
\end{itemize}

We incorporate RL into our offline streaming training. To ensure the stability and sonsistency of the training process, we only sample 1\% query per pass to run the RL policy. The final loss will be a weighted fusion of the MTP and RL loss:
\begin{equation}
    \mathcal{L}_{OneLive} := \mathcal{L}_{MTP} + w\mathcal{L}_{RL}.
\end{equation}

\section{Experiments}

In this section, we present extensive offline experiments to demonstrate the performance improvement of OneLive and its related components compared to baselines in live-streaming scenarios. We also conduct rigorous A/B tests to verify the effectiveness of OneLive in real-world online deployment. In addition, we provide case studies to offer further analysis of the gain sources of OneLive. Details of experiment settings refer to Appendix \ref{app:settings}.

\subsection{Offline Experiment}

\subsubsection{\textbf{Overall Performance.}} 

As results shown in Table \ref{tab:offline}, our proposed model OneLive consistently outperforms all existing baselines, with the maximum scale improvements on click records reaching up to $16.70\%$ in HR@128 and $14.26\%$ in MRR@128. 
It also surpasses the best baseline by at least $13.29\%$ across other indicators.
These results strongly validate the effectiveness of our generative recommendation framework with dynamic awareness in live-streaming recommendation, rather than simply applying the existing generative methods (\textit{e.g.} OneRec) to switch scenarios. 
OneLive not only rivals but even exceeds carefully engineered industrial architectures based on ANN retrieval (\textit{e.g.} KuaiFormer and GNN), thereby offering superior support for real-time content understanding and interaction prediction.
The detailed analysis of the overall experiments is presented in Appendix \ref{app:overall}.

\subsubsection{\textbf{Component Ablation.}} 

\begin{table}[ht!]
  \caption{The ablation experiments conducted in component stacking under click records. The best performances of each component are boldfaced.}
  \vspace{-0.3cm}
  \setlength{\tabcolsep}{2pt}
  \label{tab:ablation}
  \begin{tabular}{lcccc}
    \toprule

    \multicolumn{1}{c}{\textbf{Variants}} & \textbf{ACC@all}  & \textbf{Imprv. $\uparrow$} & \textbf{HR@128}& \textbf{Imprv. $\uparrow$} \\
    
    \midrule

    \textbf{OneRec} \\
    \quad \textbf{+ 3 $\times$ 512 MLLM} & 0.613 & - & 0.621 & - \\

    \midrule

    \textbf{Tokenizer} & & & & \\
    \quad + $3\times8192$ MLLM & 0.592 & -3.42\% & 0.643 & +3.54\% \\
    \quad + $3\times512$ IA & 0.616 & +0.49\% & 0.640 & +3.06\% \\
    \quad \textbf{+ 3 $\times$ 8192 IA} & 0.608 & -0.82\% & 0.684 & +10.14\% \\

    \midrule

    \textbf{Architecture} & & & & \\
    \quad + Standard Dec-Only & 0.618 & +0.82\% & 0.695 & +11.92\% \\
    \quad + Lazy Dec-Only & 0.621 & +1.31\% & 0.710 & +14.33\% \\
    \quad \textbf{+ MTP} & 0.619 & +0.98\% & 0.708 & +14.01\% \\

    \midrule

    \textbf{Temporal} & & & & \\
    \quad + Gated-Attn & 0.619 & +0.98\% & 0.710 & +14.33\% \\
    \quad \textbf{+ TA Gated-Attn} & 0.630 & +2.77\% & 0.717 & +15.46\% \\

    \midrule

    \textbf{OneLive} \\
    \quad \textbf{+ Full} & 0.641 & +4.57\% & 0.725 & +16.75\% \\

    \bottomrule
  \end{tabular}
  \vspace{-0.3cm}
\end{table}

We conduct component ablation and replacement experiments on three core modules of OneLive. 
We regard OneRec as our baseline, which adopts an Encoder-Decoder architecture to predict $3 \times 512$ MLLM code to adapt live-streaming scenario.
We incrementally stack each component, and report the metrics ACC@all of training and HR@128 of inference, with results shown in Table \ref{tab:ablation}. We summarize key observations as follows:
\begin{itemize}
    \item \textbf{Tokenizer}: Results on inference of multiple tokenizer demonstrate two conclusions: on the one hand, IA Code outperforms MLLM Code by incorporating behavior signals to obtain more discriminative SIDs; on the other sid, increasing the codebook size also yields performance gains. Both changes can reduce collision ratio thereby enhancing the individual expression of SIDs.
    Besides, we observe the misalignment between training and inference metrics, which might due to coupled data simplifies the model's learning but results in degraded accuracy and generalization during inference.
    \item \textbf{Architecture}: Results of architecture show that, under similar parameter quantity, Decoder-Only allocates more computation to the decoding process compared to Encoder-Decoder architecture, thereby enhancing the model's generation capability.
    Furthermore, the similar implementation of Lazy Decoder-Only improves upon Naive Decoder-Only by introducing explicit sequence crossover, enabling adaptive context aggregation during generation.
    The introduction of the Sequential MTP mechanism maintains comparable generation performance, 
    while achieving substantial inference acceleration, characterized by a \textbf{31.8\%} increase in single-machine QPS and a \textbf{55.45\%} reduction in computational graph execution latency, as the results shown in Table \ref{tab:mtp}. 
    Our hardware utilization is also substantially improved, achieving a MFU of \textbf{22.78\%} on the L20 GPU (compared to only \textbf{2.56\%} for the online ranking model).
    \item \textbf{Temporal}: Results of temporal modeling show that simple gated attention does not bring obvious benefits. However, after incorporating temporal side information into sequence modeling and prompt design, the model achieves adaptive attention gating based on temporal awareness.
    This leads to reasonable distribution of living authors, as the improvement of the validity rate by \textbf{14.29\%} during inference.
\end{itemize}

\subsubsection{\textbf{RL Ablation.}} 

\begin{table}[t!]
  \caption{The comparison of QPS improvement and computational graph execution latency.}
  \label{tab:ab}
  \begin{tabular}{c|lccc}
    \toprule
    \multicolumn{1}{c}{\makecell{\textbf{Number of} \\ \textbf{Layers}}} & \multicolumn{1}{c}{\textbf{Variants}} & \makecell{QPS \\Imprv. $\uparrow$} & \makecell{Latency \\(ms, Avg)} & \makecell{Latency \\(ms, P99)} \\
    
    \midrule
    
    \multirow{2}{*}{\textbf{6}} & Lazy Dec-Only & - & 6.82 & 13.34 \\
    & \quad + MTP & +62.0\% & 2.97 & 5.04 \\

    \midrule
    
    \multirow{2}{*}{\textbf{3}} & Lazy Dec-Only & - & 5.05 & 8.48 \\
    & \quad +MTP & +31.8\% & 2.25 & 3.62 \\
    \bottomrule
  \end{tabular}
  \label{tab:mtp}
\end{table}

\begin{table}[t!]
  \caption{The ablation experiment of reinforcement learning.}
  \vspace{-0.3cm}
  \label{tab:rl}
  \begin{tabular}{c|lcccc}
    \toprule

    \multicolumn{1}{c}{\multirow{2.5}{*}{\textbf{Beam Top}}} & \multicolumn{1}{c}{\multirow{2.5}{*}{\textbf{Model}}} & \multicolumn{4}{c}{\textbf{Reward}} \\
    
    \cmidrule(r){3-6} \multicolumn{1}{c}{ } & \multicolumn{1}{c}{ } & \textbf{LVTR} & \textbf{CTR} & \textbf{WTR} & \textbf{GTR} \\
    
    \midrule

    \multirow{3}{*}{\textbf{Top64}} & OneLive & 0.03808 & 0.04883 & 0.00536 & 0.00127 \\
    & + DPO & 0.03680 & 0.04723 & 0.00513 & 0.00123 \\
    & + GRPO & 0.04070 & 0.05042 & 0.00549 & 0.00136 \\

    \midrule

    \multirow{3}{*}{\textbf{Top256}} & OneLive & 0.03066 & 0.04092 & 0.00473 & 0.00113 \\
    & + DPO & 0.03131 & 0.04142 & 0.00468 & 0.00114 \\
    & + GRPO & 0.03405 & 0.04346 & 0.00494 & 0.00118 \\

    \bottomrule
  \end{tabular}
  \vspace{-0.3cm}
\end{table}

We present the ablation experiments of reinforcement learning based on OneLive in Table \ref{tab:rl}. The results show that compared with the baseline, GRPO achieves consistent performance and system recognition improvements across different beam size settings, while DPO only filters better candidates that align with user preferences when the beam size is large. 
This might because that GRPO makes use of all the responses within the group for advantages calculation and gradient optimization, which provides richer and more robust feedback signals. 
However, DPO optimizes directly based on limited feedback from positive and negative sample pairs, which increases the risk of being affected by noise in user preferences under small baem size.

\subsubsection{\textbf{Parameter Scaling.}}

\begin{figure}[t!]
\begin{center}
\includegraphics[width=8cm]{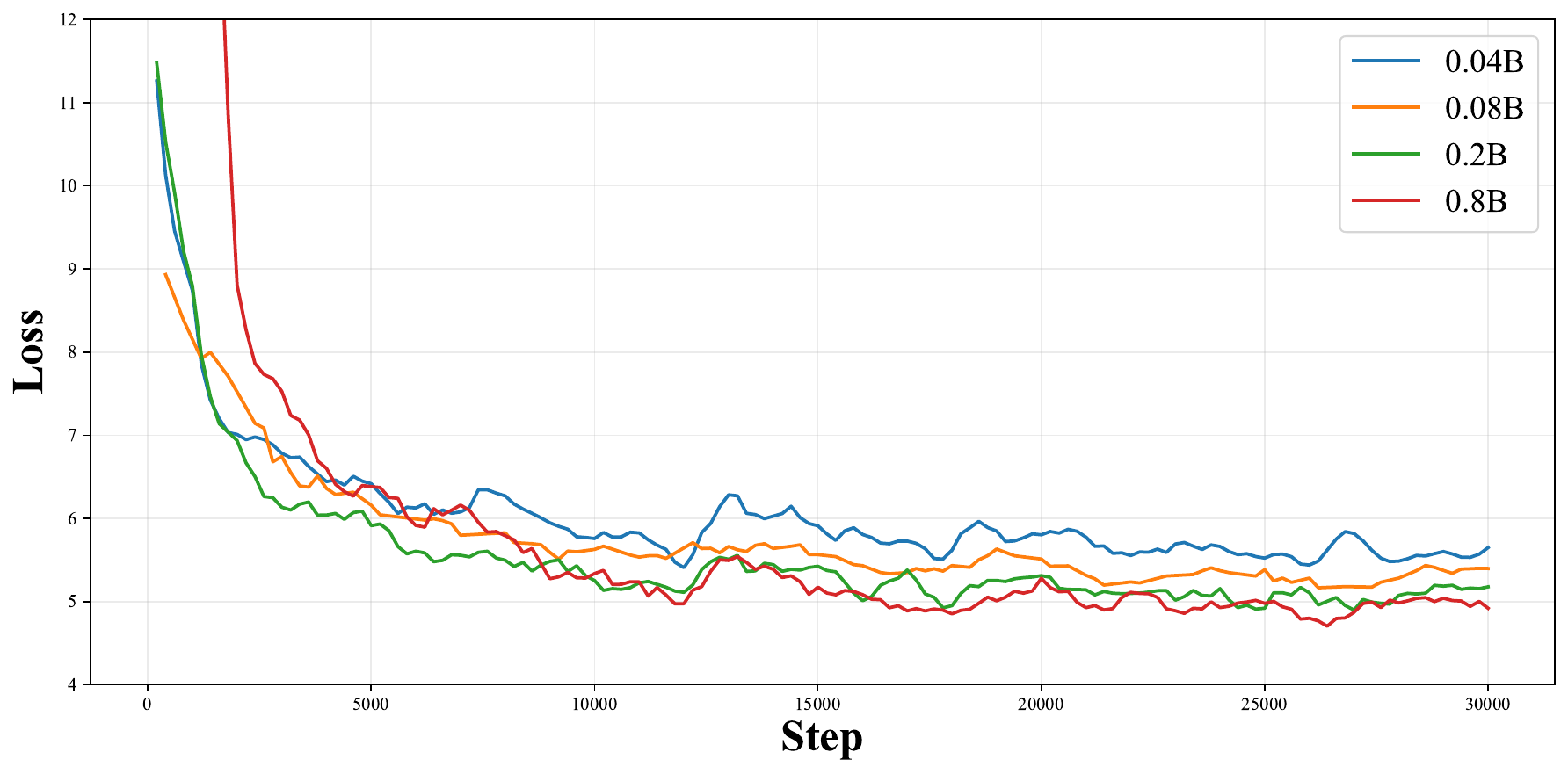}
\vspace{-0.3cm}
\caption{Loss curves under parameter scaling.}
\vspace{-0.5cm}
\label{fig:scaling}
\end{center}
\end{figure}

To investigate the suitability of scaling law in live-streaming generative recommendation model, we conduct a model scaling experiment. We train model across different parameter scales and visualize the loss curves in Figure \ref{fig:scaling}. The results exhibit that, as model size increases, the performance shows a stable upward trend, with loss consistently decreases. 
However, the gains exhibit a diminishing marginal returns effect, as the loss decrease brought by further scaling based on larger parameter scale gradually slows down.
Considering the online deployment constraints, we finally deploy 0.08B version for online service.

\subsection{Online A/B Test}

\begin{table}[t!]
  \caption{Online A/B test performance in live-streaming services of  Kuaishou and Kuaishou Lite.}
  \vspace{-0.3cm}
  \setlength{\tabcolsep}{1.5pt}
  \label{tab:ab}
  \begin{tabular}{cccccc}
    \toprule

    \textbf{Application} & Exposure & CTR & Click Count & Watch Time & Follow \\
    
    \midrule

    \textbf{Kuaishou} & +1.32\% & +0.41\% & +1.73\% & +0.58\% & +1.36\% \\
    \textbf{Kuaishou Lite} & +1.96\% & +0.72\% & +2.70\% & +0.41\% & +2.07\% \\
     
    \bottomrule
  \end{tabular}
  \vspace{-0.3cm}
\end{table}

We deploy OneLive on Kuaishou and Kuaishou Lite, providing servies to hundreds of millions of users and millions of authors on two major live-streaming platforms, to observe the real effectiveness of our model's performance for online business.
We launch our optimal combo model version to the production environment, and conduct continuous online observation for nearly one month. The results are summarized in Table \ref{tab:ab}.

OneLive achieves significant gains in user penetration within the live-streaming recommendation scenario. 
Across two real world applications, both exposure and CTR show substantial improvements at the page level, which validates the significant advantage of our unified generative architecture over the online recommendation pipeline. 
Besides, multiple behavior indicators such as click, watch and follow, also increase noticeably.
These results demonstrate that our model accurately captures user preferences and enables efficient content distribution. 

\section{Conclusion}

In this paper, we propose dynamically unified generative framework OneLive tailored for live-streaming recommendation.
We capture and quantify live-streaming real-time content using a dynamic tokenizer, perform temporal modeling with time-aware gated attention, achieve stable training and optimized inference through Sequential MTP and QK normalization, and align personalized preferences via multi-objective reinforcement learning.
Extensive offline and online experiments demonstrate the effectiveness of OneLive.
Now OneLive has been deployed on Kuaishou App and serving 400 million users daily in live-streaming scenario, bringing significant business benefits.
In the future, we will continue to explore the more effective and efficient live-streaming generative recommendation.


\balance
\bibliographystyle{ACM-Reference-Format}
\bibliography{sample-base-extend.bib}


\appendix

\section{Related Work}

\subsection{Live-streaming Recommendation}

Live-streaming is an emerging popular media that allows for direct interaction between users and authors, which has been widely explored on many online platforms, such as Kuaishou~\cite{moment} and TikTok~\cite{tsstfn}. Compared with short videos or products, the most significant difference of live-streaming lies in the dynamic changes of its content. Therefore, live-streaming recommendation have strict requirements for real-time content understanding and fast response. 
Early methods such as LiveRec~\cite{liverec} only focus on historical interactions and fail to incorporate content information.
ContentCTR~\cite{contentctr} and MMBee~\cite{mmbee} introduce multimodal information about the granularity of the entire live-streaming and develop a more comprehensive understanding, but such rough aggregated information may not match the real-time state. 

In order to achieve dynamic perception of live-streaming, Moment~\cite{moment} changes the data-streaming engine to a real-time segment report manner. On this basis, either FARM~\cite{farm} introduces cross-domain knowledge or LARM~\cite{larm} adds LLM encoding, both of which fully enhance the ability of modeling relationship between users and authors. 
However, these discriminative methods still rely on pointwise scoring and ranking, and are limited by the pre-filtering of the candidate set, making end-to-end prediction at scale challenging.
LiveForesighter~\cite{liveforesighter, foresight} generates predictions for the next product in online-shopping live-streaming, but the results are still incorporated into a feature cross-fusion module, without changing the essence of the recommendation paradigm. 

Therefore, shifting from a cascaded discriminative architecture to an end-to-end generative paradigm is crucial to meet the demand for immediate feedback in live-streaming recommendation.

\subsection{Generative Recommendation}

Generative models, especially LLMs~\cite{qwen3, deepseek-r1}, have achieved breakthrough progress in various fields and demonstrated great potential in recommendation systems. 
The main research line of LLM-based generative recommendation is to generate personalized suggestions through prompts or fine-tuning using pretrained LLMs~\cite{gr-survey}. These methods either leverage contextual awareness~\cite{lc-rec, eager-llm} or stimulate reasoning capability~\cite{reason4rec, rearec} of LLMs to align recommendation tasks, but often require additional data restructuring or time consumption.

Another insight from LLMs lies in the benefits of scaling, which has been proven the effectiveness within the cascaded stages~\cite{hstu, rankmixer}. 
To achieve scaling in recommendation and overcome the computational fragmentation caused by cascaded architectures, 
directly training a real end-to-end large recommendation model has attracted attentions from the industry. TIGER~\cite{tiger} establishes the foundational architecture for this paradigm, which represents items as sequence of discrete tokens from codebooks learned by RQ-VAE~\cite{rq-vae}, then uses encoder-decoder model to directly predict the Semantic ID of the next item.
OneRec~\cite{onerec} implements this end-to-end architecture in industry, replacing the cascaded pipeline, and introduces preference alignment to optimize generation results. On this basis, OneLoc~\cite{oneloc} and OneSug~\cite{onesug} adapt this idea in the scenarios of local life and query suggestion. 
COBRA~\cite{cobra} unifies the generation of sparse and dense, EGA-V2~\cite{ega-v2} integrates multiple tasks, and MMQ~\cite{mmq-v2} aligns content and behavior for Semantic ID learning. These methods have significantly advanced this direction, yet remain underexplored in live-streaming scenario.

\section{Experiment Details}

\subsection{Experimental Settings} \label{app:settings}

\subsubsection{\textbf{Dataset.}}

We construct training and testing datasets based on large-scale industrial live-streaming platform logs from the Kuaishou App. We collect data for nearly two months, including 400 million users and 3 million authors, as well as billions of interaction data between them, covering long-view and click records. The data from the last day are used for testing, and the data from the remaining dates are used for training. All subsequent offline and ablation experiments are conducted on this dataset.

\subsubsection{\textbf{Evaluation Metrics.}}

Following most prior works, the evaluation considers both recall and ranking performance, therefore we adopt Accuracy (ACC), Hit Rate (HR) and Mean Reciprocal Ranking (MRR) as our offline evaluation metrics. 
We also utilize Reward evaluated by the deployed online ranking model to quantify its recognition of the generative results.
For online A/B tests, we use core market indicators to demonstrate the effectiveness to online business. Detailed metric definitions are provide in Appendix \ref{app:metrics}.

\subsubsection{\textbf{Baselines.}}

We compare OneLive with competitive baselines within two groups of works: (i) \textbf{Traditional Models}: classic sequential recommendation method \textbf{SASRec}~\cite{sasrec}, adaptive long sequence compression method \textbf{KuaiFormer}~\cite{kuaiformer} and heterogeneous structure-aware graph method \textbf{GNN}~\cite{lightgcn} variant; (ii) \textbf{Generative Models}: original SID-based generative method \textbf{TIGER}~\cite{tiger} and successful solution in industrial scenario \textbf{OneRec}~\cite{onerec}.

\subsection{Overall Experiemnt Analysis} \label{app:overall}

Notably, not all generative methods consistently outperform traditional methods. In particular, TIGER shows a significant performance drop, which contradicts with prior industrial reports~\cite{onerec, oneloc}. 
We attribute this to the fact that in the dynamic scenario of live-streaming, TIGER models the SID sequence corresponding to the user's historical interaction items (MLLM Codes of authors in reproduction experiment, while even worse with SID Codes for capturing instant rather than synthesized content) for subsequent prediction, which hinders the model's ability to distribute dynamically from such coarse-gained and static representations. 
This highlights and indirectly confirms the core challenge of providing real-time recommendation with evolving content in live-streaming scenario.

However, OneLive and OneRec demonstrate strong adaptability in industrial settings. Beyond the discrete feature SID, the incorporation of extensive user and author side information significantly enhances the model's generalization capability in complex real-world environments. 
Moreover, the introduction of 
temporal modeling and enriched training objectives
better aligns with the requirements of live-streaming recommendation.

\subsection{Online Stratified Test}

\begin{figure}[t!]
    \centering
    \subfloat[Kuaishou\label{fig:group_main}]{
        \includegraphics[width=8.5cm]{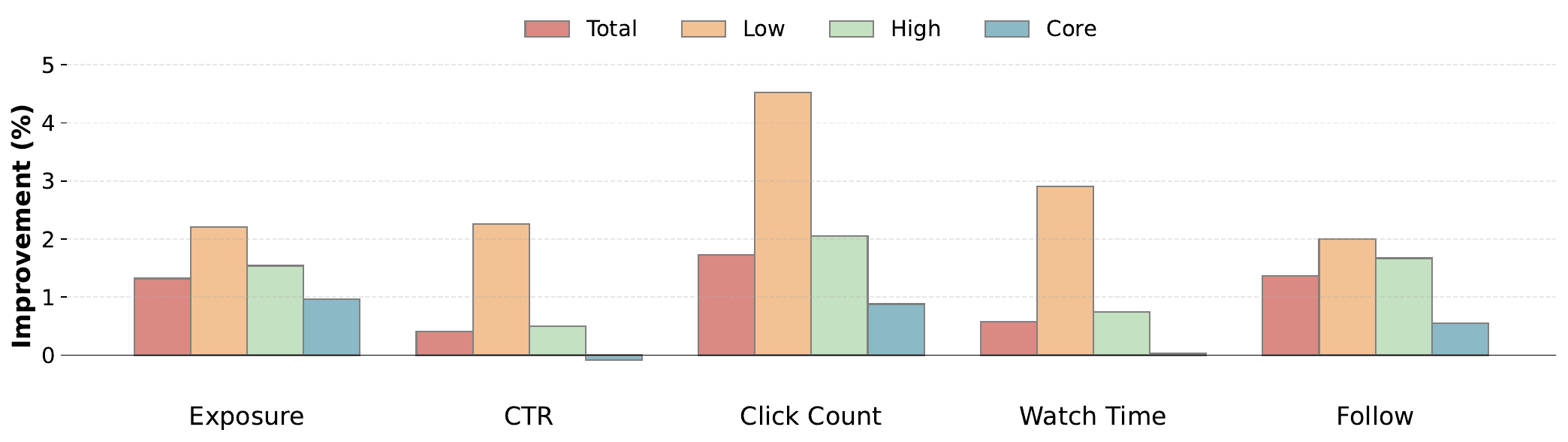}
    }
    \quad
    \subfloat[Kuaishou Lite\label{fig:group_lite}]{
        \includegraphics[width=8.5cm]{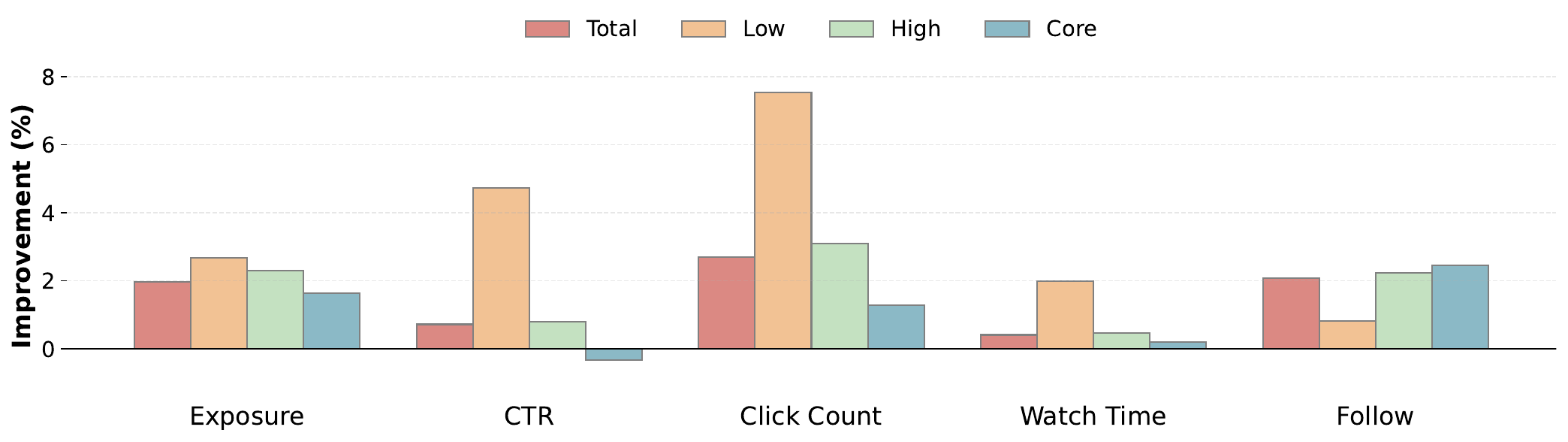}
    }
    \caption{Stratified test of online performance.}
    \label{fig:group}
\end{figure}

We further conduct a stratified test to evaluate the generalization over distinct crowd groups, including low-active users, high-active users and core-paid users. The performance across these groups are visualized in Figure \ref{fig:group}.
Except for minor decrease in a few indicators, the results show that our unified generative model achieves consistent and significant improvements across all groups.
Notably, low-active users exhibit the largest gains, which might be attributed to the long-tail neglect problem in conventional cascade pipelines.
In contrast, OneLive demonstrates an astonishing potential to meet diverse  preferences across active levels.

\subsection{Case Study}

\begin{figure}[t!]
    \centering
    \subfloat[Talent Show Author.\label{fig:case_1}]{
        \includegraphics[width=8.5cm]{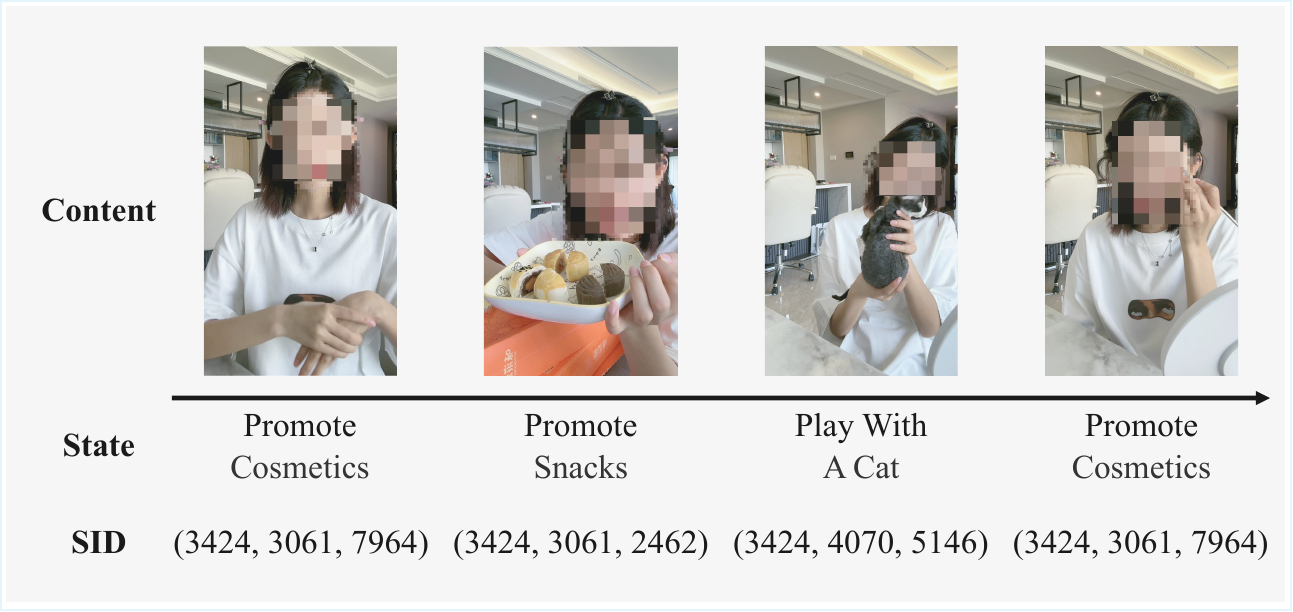}
    }
    \quad
    \subfloat[E-Commerce Author.\label{fig:case_2}]{
        \includegraphics[width=8.5cm]{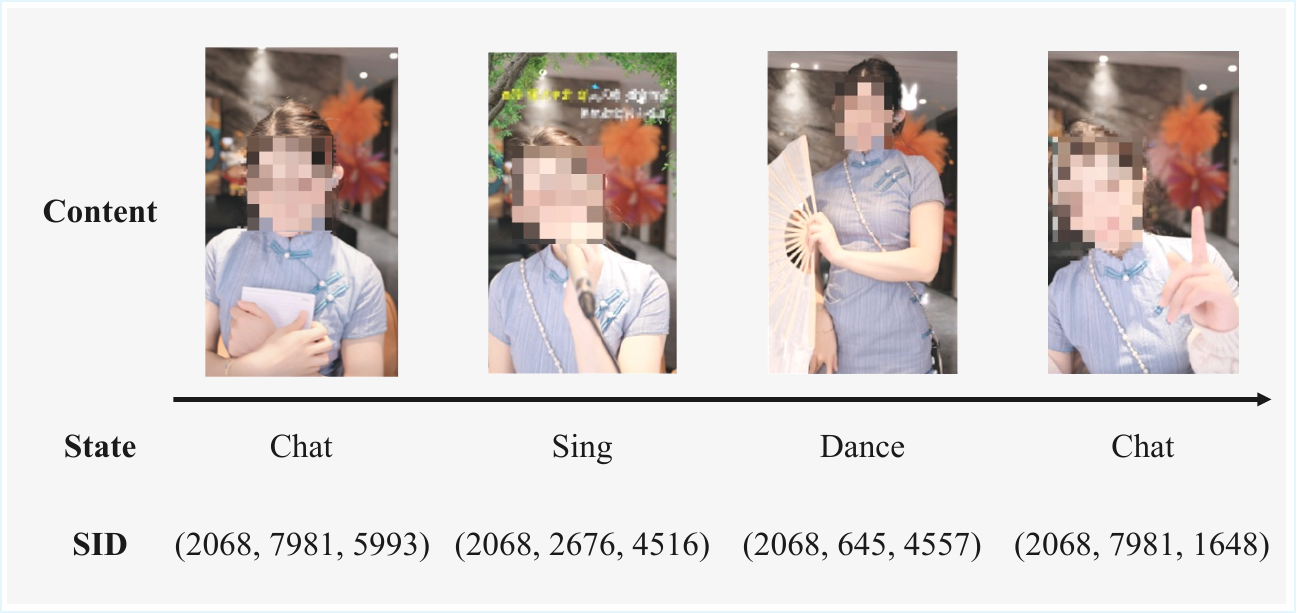}
    }
    \caption{Cases of SID updates with live-streaming evolving.}
    \label{fig:case}
\end{figure}

We demonstrate several cases in Figure \ref{fig:case}. We can observe that, our live-streaming SID generated from dynamic tokenizer can capture the current live content of authors and update in real time, ensuring continuous continuity with live-streaming evolving.

\section{Metrics Definitions}

\subsection{Codebook Analysis} \label{app:codebook}

We use \textbf{Utilization Rate (UR)} and \textbf{Collision Rate (CR)} to evaluate the quality of SID code. Specifically:
\begin{itemize}
    \item \textbf{Utilization Rate (UR)}: It measures the ratio of the number of distinct codes appearing at each layer relative to the size of the entire codebook size. For a given layer $L_i$, let $C_{total}$ represent the total size of the codebook, and $C_{used}$ denote the number of distinct codes that have appeared at this layer. The $\mathbf{UR}_{L_i}$ can be expressed as:
    \begin{equation}
        \mathbf{UR}_{L_i} = \frac{C_{used}}{C_{total}}.
    \end{equation}
    \item \textit{Collision Rate (CR)}: This measures the degree of overlap from different perspectives (\textit{e.g.}, SID or Author). \\
    Specifically, in the SID dimension, a SID may be associated with multiple different authors. Therefore, the $\mathbf{CR}_{SID}$ measures the ratio of the number of SIDs $C_{collision\_sid}$ that are associated with multiple authors relative to the total number of distinct SIDs $C_{sid}$:
    \begin{equation}
        \mathbf{CR}_{SID} = \frac{C_{collision\_sid}}{C_{sid}}.
    \end{equation}
    \\
    In the Author dimension, multiple authors may share the same SID. Therefore, the $\mathbf{CR}_{Author}$ measures the ratio of the number of authors $C_{collision\_author}$ who share same SID with others relative to the total number of distinct authors $C_{author}$:
    \begin{equation}
        \mathbf{CR}_{Author} = \frac{C_{collision\_author}}{C_{author}}.
    \end{equation}
\end{itemize}

\subsection{Experiment Comparison} \label{app:metrics}

We use \textbf{Accuracy (ACC)} as the metric for training, and \textbf{Hit Rate (HR)}, \textbf{Mean Reciprocal Ranking (MRR)} and \textbf{Reward} for inference. Specifically:
\begin{itemize}
    \item \textbf{Accuracy (ACC)}: This measures the average accuracy of multi-level SID hits during training. Formally:
    \begin{equation}
        \mathbf{ACC}@all = \frac{1}{N}\sum^N_{i=1}\mathbb{I}({\hat{q}}_{i}=q_{i}) ,
    \end{equation}
    where $N$ is the total SID in training samples, ${\hat{q}}_{i}$ is the predicted SID and $q_{i}$ is the true SID.
    \item \textbf{Hit Rate (HR)}: This measures the correct answer appears in the Top-k predictions when performing inference with beam search. Formally:
    \begin{equation}
        \mathbf{HR}@k = \frac{1}{N}\sum^N_{i=1}\mathbb{I}(y_i\in \text{Top-k}),
    \end{equation}
    where $y_i$ is the ground truth and $\text{Top-k}$ refers to the predictions generated by the model.
    \item \textbf{Mean Reciprocal Ranking (MRR)}: This measures the rank of the correct label within the Top-k predictions. Formally:
    \begin{equation}
        \mathbf{MRR}@k = \frac{1}{N}\sum^N_{i=1}\frac{1}{\text{rank}_i},
    \end{equation}
    where $\text{rank}_i$ is the position of the correct label in the Top-k predictions.
    \item \textbf{Reward}: This measures the recognization of online system based on how well it ranks the predictions. It obtains the XTR scores by transferring the user and author pairs obtained from the prediction results into the ranking model. Formally:
    \begin{equation}
        \mathbf{Reward}_{XTR} = \frac{1}{N}\sum^N_{i=1}\text{Ranking}_{XTR}(\text{User}, y_i).
    \end{equation}
\end{itemize}

\end{document}